\documentclass[%
 reprint,
superscriptaddress,
 amsmath,
 amssymb,
 prl,
]{revtex4-2}

\usepackage{
amsmath,
amssymb,
bm,
color,
dcolumn,
enumerate,
etoolbox,
float,
graphicx,
hyperref,
soul,
tabularx, 
titlesec,
xspace,
todonotes,
}

\titleformat{\section}
{\sffamily\bfseries}
{\thesection}{0em}{\MakeUppercase}
\titleformat{\subsection}
{\small\sffamily\bfseries}
{\thesection}{0em}{}
\titlespacing\section{0pt}{8pt}{0pt}
\titlespacing\subsection{0pt}{0pt}{0pt}

\emergencystretch=\maxdimen
\definecolor{linkColor}{rgb}{1,0,0}
\hypersetup{pdfborder={0 0 0},colorlinks=true,urlcolor=linkColor,citecolor=linkColor}

\newcommand{\pyFDSTEM}{\href{https://github.com/py4dstem/py4DSTEM}{\texttt{py4DSTEM}}}
\newcommand{\prismatic}{\href{https://prism-em.com/}{\texttt{Prismatic}}}
\newcommand{\crystalFD}{\href{https://github.com/AI-ML-4DSTEM/crystal4D/tree/dev}{\texttt{crystal4D}}}

\begin{document}

\title{Disentangling multiple scattering with deep learning: application to strain mapping from electron diffraction patterns}

\author{Joydeep Munshi}
\affiliation{Center for Nanoscale Materials, Argonne National Laboratory, Lemont, IL, USA  60517}

\author{Alexander Rakowski}
\affiliation{National Center for Electron Microscopy, Molecular Foundry, Lawrence Berkeley National Laboratory, Berkeley, CA, USA 94720}

\author{Benjamin H Savitzky}
\affiliation{National Center for Electron Microscopy, Molecular Foundry, Lawrence Berkeley National Laboratory, Berkeley, CA, USA 94720}

\author{Steven E Zeltmann}
\affiliation{Department of Materials Science and Engineering, University of California, Berkeley, CA,  94720}

\author{Jim Ciston}
\affiliation{National Center for Electron Microscopy, Molecular Foundry, Lawrence Berkeley National Laboratory, Berkeley, CA, USA 94720}

\author{Matthew Henderson}
\affiliation{Scientific Data Division, Lawrence Berkeley National Laboratory, Berkeley, CA, USA 94720}

\author{Shreyas Cholia}
\affiliation{Scientific Data Division, Lawrence Berkeley National Laboratory, Berkeley, CA, USA 94720}

\author{Andrew M Minor}
\affiliation{National Center for Electron Microscopy, Molecular Foundry, Lawrence Berkeley National Laboratory, Berkeley, CA, USA 94720}
\affiliation{Department of Materials Science and Engineering, University of California, Berkeley, CA,  94720}

\author{Maria K Y Chan}
\email{mchan@anl.gov}
\affiliation{Center for Nanoscale Materials, Argonne National Laboratory, Lemont, IL, USA  60517}

\author{Colin Ophus}
\email{cophus@gmail.com}
\affiliation{National Center for Electron Microscopy, Molecular Foundry, Lawrence Berkeley National Laboratory, Berkeley, CA, USA 94720}

\date{\today}
\begin{abstract}


Implementation of a fast, robust, and fully-automated pipeline for crystal structure determination and underlying strain mapping for crystalline materials is important for many technological applications. Scanning electron nanodiffraction offers a procedure for identifying and collecting strain maps with good accuracy and high spatial resolutions. However, the application of this technique is limited, particularly in thick samples where the electron beam can undergo multiple scattering, which introduces signal nonlinearities. Deep learning methods have the potential to invert these complex signals, but previous implementations are often trained only on specific crystal systems or a small subset of the crystal structure and microscope parameter phase space. In this study, we implement a Fourier space, complex-valued deep neural network called FCU-Net, to invert highly nonlinear electron diffraction patterns into the corresponding quantitative structure factor images. We trained the FCU-Net using over 200,000 unique simulated dynamical diffraction patterns which include many different combinations of crystal structures, orientations, thicknesses, microscope parameters, and common experimental artifacts. We evaluated the trained FCU-Net model against simulated and experimental 4D-STEM diffraction datasets, where it substantially out-performs conventional analysis methods. Our simulated diffraction pattern library, implementation of FCU-Net, and trained model weights are freely available in open source repositories, and can be adapted to many different diffraction measurement problems.

\end{abstract}
\maketitle

\section*{Introduction}
\label{section:intro}

Scanning transmission electron microscopy (STEM) has emerged as one of the primary nanoscale materials characterization tools \citep{liu2021advances}. A STEM experiment focuses an electron beam on to a sample, with the probe dimensions ranging from tens of nanometers down to the atomic scale, which is made possible by hardware aberration correction \citep{haider1998electron, krivanek2003towards}. STEM experiments have successfully measured the 2D position of atomic columns with picometer-precision \citep{yankovich2014picometre}, measured the vibrational spectra of single-atom defects \citep{hage2020single}, mapped solid-liquid interfaces in lithium-metal batteries \cite{zachman2018cryo}, and determined the 3D position and chemical species of each atom in a nanoparticle \cite{yang2017deciphering}. Atomic-resolution STEM methods provide extremely high resolution for both spatial and spectroscopic mapping, but have a limited field of view (FOV) because of the necessary minimum sampling rate required to resolve atoms \cite{yankovich2015high}. 

An alternative to real space imaging in STEM is to instead record a converged beam electron diffraction (CBED) pattern at each probe position, resulting in a four dimensional (4D-STEM) dataset \citep{ophus2019four}. 4D-STEM experiments are gaining popularity among electron microscopists because they can collect atomic-scale information from each probe over a nearly arbitrary field-of-view \cite{ozdol2015strain}, and can measure a broad spectrum of quantities of physical interest including 3D structural determination \citep{nord2019three}, ferroelectric polarization \cite{das2019observation}, imaging of lithium in cathode materials \citep{ahmed2020visualization}, ptychographic atomic imaging \citep{chen2021electron}, correlation of local strain with composition from x-ray ptychography \citep{hughes2021correlative}, distinguishing between chemical and structural interfacial roughness \citep{oxley2020deep}, strain in 2D material bilayers \citep{kazmierczak2021strain, zachman2021interferometric}, and many others.  The ability to extract quantitative information with atomic-scale resolution is, however, frequently limited by the size and complexity of experimental 4D-STEM data.  Open source computational tools such as pyxem in hyperSpy \citep{duncan_n_johnstone_2021_4436723}, liberTEM \citep{clausen2020libertem}, and py4DSTEM \cite{savitzky2021py4dstem} provide high-throughput multimodal data analysis tools to the community.


Computational analysis of diffraction images from crystalline materials typically begins with localizing any Bragg scattering. A standard approach to this problem is matching a template - usually an image of the electron beam over vacuum - to each diffraction pattern using cross correlation. However, the Bragg disk intensities can oscillate with changing sample thickness, bias asymmetrically due to mistilt of the crystal zone axis relative to the electron beam, form interference effects between overlapping disks, and generally display highly nonlinear signals in all but the very thinnest of samples due to dynamical/multiple scattering \cite{mahr2015theoretical, williamson2015quantitative, grieb2018strain, zeltmann2020patterned}. While the physics of these phenomena are understood and the effects may be readily recognisable to a human observer, writing classical algorithms which can accommodate them is challenging.  Various approaches have been implemented, including cross, phase, and hybrid correlations \citep{pekin2017optimizing}, edge filtering \citep{mukherjee2020lattice}, circular Hough transforms \citep{YUAN2019112837}, and radial gradient maximization \citep{muller2012scanning}.  Zeltmann \textit{et al.} fabricated patterned apertures which result in bullseye shaped electron probes that improve the precision of disk position \citep{zeltmann2020patterned} measurements.  Other authors use Fourier space methods to pool information about the disk spacing, such as the cepstral transform \cite{padgett2020exit}. In addition to the challenge of accuracy, traditional approaches often require careful parameter tuning to achieve acceptable results, and may be time consuming \cite{maclaren2021comparing}.  Moreover, the quantity one is ideally after is not just the disk positions but the structure factors $V_g$, the positions and amplitudes of which reflect the reciprocal lattice of the scattering crystal.

Once the Bragg disks have been measured, many subsequent analyses become possible, including crystallographic orientation mapping, off-axis virtual imaging modalities, and mapping the local strain \cite{seyring2011advance, shukla2016study, pekin2017optimizing, ophus2019four}. Spatially-resolved strain maps of crystalline and semi-crystalline materials systems are important in various engineering and technological applications. For instance, local strain distortions can play an important role in tuning electronic properties of semiconductors \citep{bedell2014strain, chidambaram2006fundamentals}, and lattice deformation and distortions due to defects and doping can be characterized from localized strain maps in metals \citep{wang2012sample, zhang2013strain, chen2021bending}.

Artificial intelligence and machine learning (AI/ML) algorithms are increasingly being implemented in materials characterization, including in electron microscopy \citep{Ede_2021}. Deep learning approaches have been been demonstrated to outperform classical algorithms in variety of computer vision problems in microscopy including classification and segmentation problems \citep{George2020CASSPERAS, roberts2019deep, Ziatdinov2019}. For instance, deep convolutional neural networks (CNNs) are implemented in the analysis of images collected with various microscopy techniques such as crystal phase classification from back-scattered diffraction patterns \cite{Kaufmann2020}, structure measurement from electron diffraction and atomic-resolution STEM images \cite{Aguiar2019} and from scanning tunneling microscopy \cite{vasudevan2018mapping}, crystal symmetry identification from X-ray diffraction \cite{tiong2020identification}, defect analysis from atomic-resolution STEM images \cite{ChiaHao2020DLStrainMappingDefects}, crystal tilt and thickness detection from position averaged CBED patterns \cite{zhang2020atomic,xu2018deep}, and orientation and strain mapping from 4D-STEM diffraction datasets \cite{li2019manifold, yuan2021training}. Li \textit{et al.} used manifold learning to directly classify different features in 4D-STEM data \cite{li2019manifold}. Recently, Yuan \textit{et al.} demonstrated the possibility of using CNNs to predict high precision orientation and strain maps of crystalline systems using 4D-STEM data, computing strain in field effect transistors with both a CNN and a more traditional Hough transform approach \cite{yuan2021training}.  This work has shown the potential of supervised learning in 4D-STEM analysis and motivated towards achieving automated analysis of massive 4D diffraction dataset. 


Bragg disk position and the underlying strain field measurement of crystalline and semi-crystalline samples, leveraging supervised machine learning, can be considered as pixel-wise mapping of diffracted disk intensities to the underlying structure factors. Such tasks may be accomplished, for example, by a traditional U-Net architecture consisting of symmetric contracting (encoder) and expansive (decoder) paths, with the crucial addition of skip layer connections enabling the flow of localized contextual information from low resolution encoded features to higher resolution upsampled layers \citep{ronneberger2015unet}. 
However, while the U-Net seems to be a prudent choice for the Bragg disk measurement problem, using traditional 2D convolutional layers for the network building blocks poses a challenge: for identical samples, changing microscope parameters such as the probe semiangle will substantially change the measured diffraction images. We require a method to encode these changing experimental parameters into the signal inversion, which is not possible in the original U-Net architecture. Additionally, small shifts of the disks can be measured  using cross correlation of a probe template, but this signal is most accurately measured as the phase component of the complex-valued Fourier transform of the correlation. To preserve all the relevant signal including the complex phase, we implement a modified U-Net architecture using fully complex 2D convolutional blocks. Historically, complex representations of images and signals have numerous advantages and outperform their non-complex equivalent forms \cite{danihelka2016associative,arjovsky2016unitary, wisdom2016full, sampat2009complex}. 



\begin{figure*}[htbp]
    \includegraphics[width=1.0\textwidth]{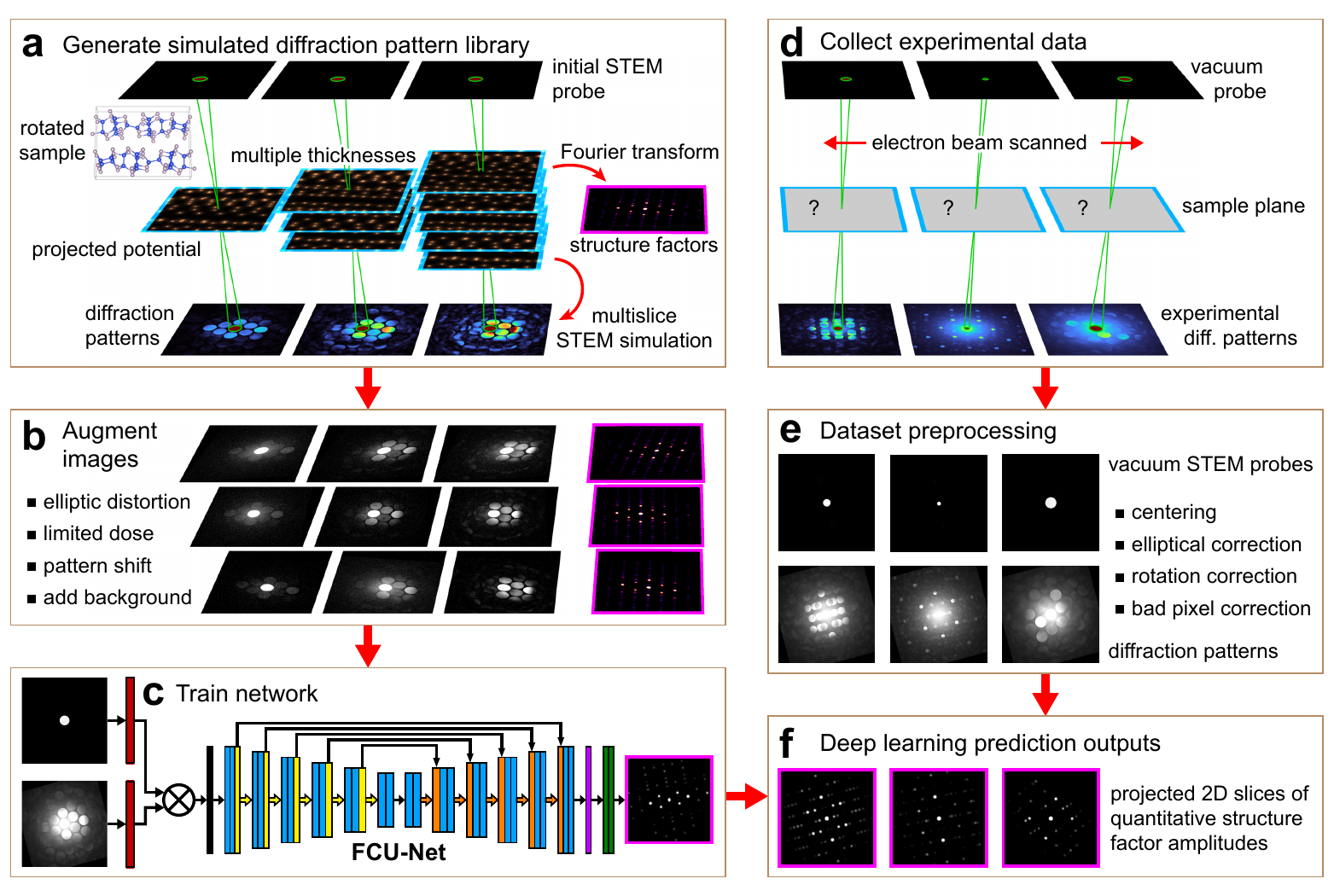}
    \caption{{\bf Overview of the methods used in this paper.} (a) Multislice diffraction simulations of many samples with different crystal structures, compositions, orientations, and thicknesses, using various microscope parameters. (b) Augmentation of the simulated images by applying elliptic distortion, pattern shift, limited signal-to-noise, and background functions. (c) Deep learning training. (d) Experimental geometry for diffraction pattern measurements. (e) Dataset preprocessing. (f) Inversion of experimental diffraction images to predict the structure factors using the FCU-Net trained in (c).}
    \centering
    \label{Fig:overview}
\end{figure*}

The complex representation is an elegant method to preserve phase information and mimics biological behavior in neurons \cite{shi2006importance}. Rippel \textit{et al.} implemented a Fourier representation of traditional CNNs  by parameterizing convolutional kernels in spectral domain  \cite{rippel2015spectral}. 
In a recent effort, Trabelsi \textit{et al.} provided building blocks for deep complex-valued convolution networks and implemented their network on a variety of deep learning tasks such as image classification, image recognition, and music and speech transcription problems \cite{trabelsi2017deep}. Here, we extend these approaches to modify the U-Net architecture to accommodate the complex and nonlinear correlation between the CBED images and the structure factors.

In this work, we implement a Fourier-space complex U-Net (FCU-Net) deep neural network which learns the mapping from measured diffraction pattern intensities to a material's underlying structure factors (Fig. \ref{Fig:overview}). We train our network on a dataset with over 200,000 unique simulated dynamical CBED data spanning thousands of crystal systems with a variety of random zone axes, off-zone tilts, thicknesses, and microscope parameters.  The training data sets are extended with physics-informed image augmentation through the addition of a realistic background, noise, and geometric distortions of the CBED patterns.  We compare the accuracy of the FCU-Net outputs to the approach of cross correlation template matching, benchmarking against the ground truth structure factors for simulated data. We further test and compare these two methods by measuring local strain using the structure factor outputs, for both simulated and experimental diffraction data of a SiGe multilayer stack, and with experimental hexagonal boron nitride 4D-STEM data. We find that FCU-Net significantly improves the accuracy of disk detection, as well as downstream measurements such as strain.  The FCU-Net pipeline is fast, highly automated, performant on materials and microscope parameters on which it has not been trained, and is robust against both experimental error and background noise.

\section*{Results and Discussion}
\label{section:results}

\subsection*{Comparison of traditional and complex U-Nets}

To start with the disk position measurement, we implement supervised learning on a large training dataset consisting of simulated CBED images and structure factors. To map diffracted disk intensities to the structure factors, we implement three variants of CNN architecture: real-valued U-Net, a U-Net with spectral parameterization, and the fully complex variant, FCU-Net. Fig.~\ref{Fig:overview} summarizes the overview of this work, where Fig.~\ref{Fig:overview}a-c show the methods we use to train the machine learning models from the simulated STEM diffraction pattern and the underlying structure factors. Fig.~\ref{Fig:overview}d-f show the inference stage to predict structure factors from experimental diffraction patterns.
 The computational methods implemented to simulate training data, architecture of the CNN models implemented in this work, the training process, and implementation and inference from experimental diffraction patterns can be found in the Methods \ref{section:methods} section.

\begin{figure*}[htbp]
    \includegraphics[width=1.0\textwidth]{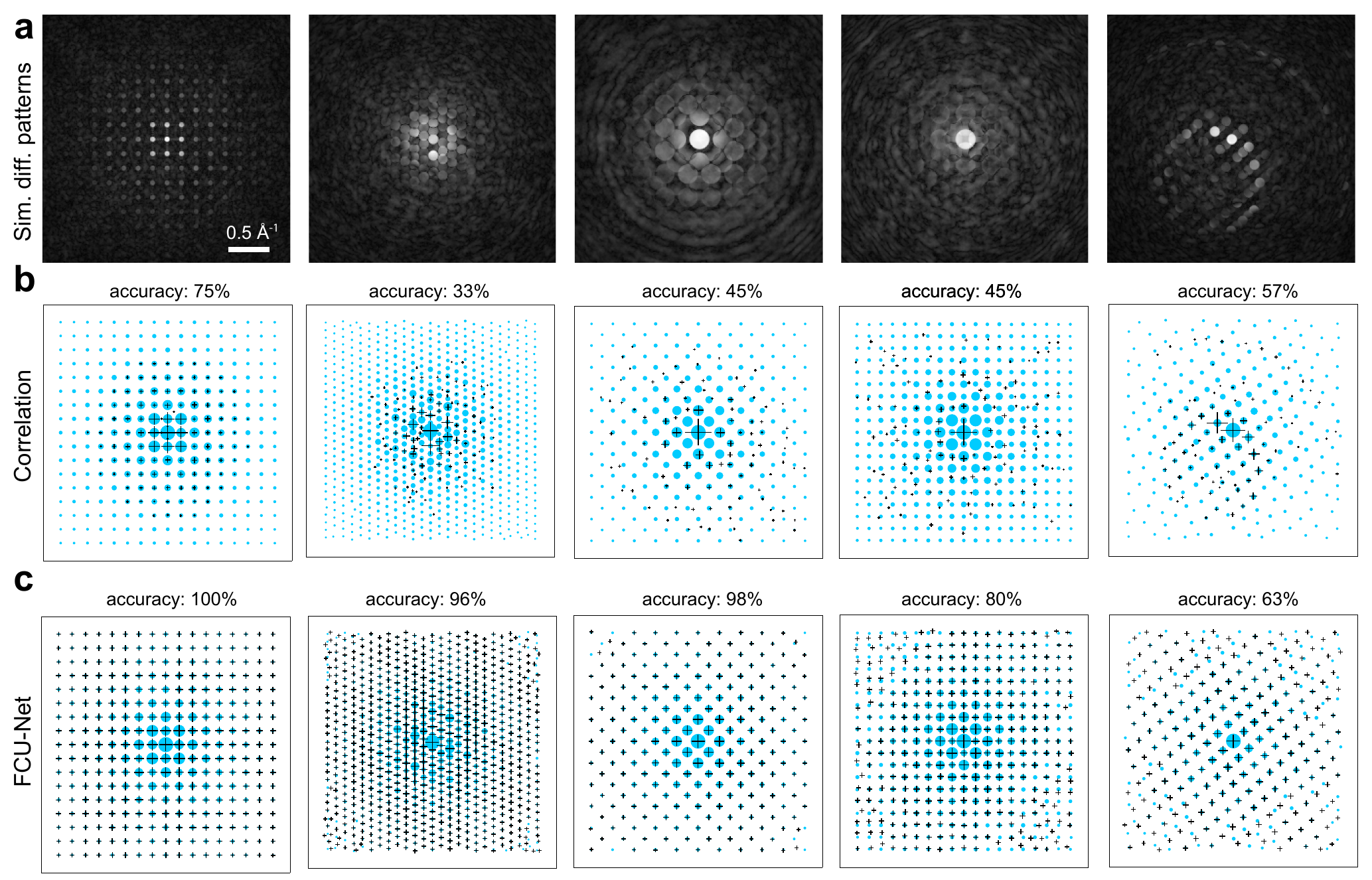}
    \caption{{\bf Bragg disk detection using cross correlation and deep learning methods.} (a) Examples of simulated diffraction patterns for crystals of different thicknesses and orientations. (b,c) The positions of the ground truth structure factor coefficients of the crystal lattice are plotted below as blue circles, with a size proportional to the structure factor amplitudes $V_g$. The structure factor positions were computed using (b) template matching by cross correlation with the vacuum probe signal, and (c) the FCU-Net network. Both measurements are overlaid as black crosses, with a size proportion to the estimated disk amplitude (square root of the disk intensity) and $V_g$ amplitudes, for the correlation and FCU-Net predictions, respectively. The total intensity-weighted accuracy is listed above for all measurements.}
    \centering
    \label{Fig:disk_compare}
\end{figure*}

Once the networks are trained, we predict the structure factors of diffraction patterns from the simulated test dataset and used them to compute the structural similarity index (SSIM), a metric of image similarity measurement \citep{wang2004image}. Table \ref{table:FCU_netCompare} compares the results for different CNN models. We find a significant improvement in the SSIM scores measured on the test dataset for the FCU-Net model, compared to networks without spectral pooling and/or without complex convolutional layers. The improvement in the overall model efficiency for the high-tilt, off-zone samples is more prominent than in the untilted, on-zone samples. We attribute this to the sensitivity of FCU-Net to the phase component of the input signal, as we expect the contribution of the phase to be more significant for high-tilt samples due to the asymmetry of their diffraction images.

\begin{table}[h!]
\caption{Accuracy of the recovered structure factor images, evaluated using the SSIM on the test dataset.}
\begin{tabular}{m{2.9cm}cc} 
 \hline
 & untilted/on-zone & high-tilt/off-zone \\ [0.5ex] 
 \hline
 U-Net (traditional) & 0.923 & 0.750  \\ 
 U-Net (spectral) & 0.926 & 0.781  \\
 FCU-Net & \textbf{0.948} & \textbf{0.880} \\
 \hline
\end{tabular}
\label{table:FCU_netCompare}
\end{table}

\subsection*{Accuracy of diffracted disk position measurements}

To evaluate the accuracy of Bragg disk detection using the trained FCU-Net and using cross correlation, we calculate the intensity weighted accuracy of the disk locations determined by each method, using the simulated test dataset with different crystal orientations and in-plane rotations. The intensity-weighted accuracy is defined as
\begin{equation}
    \rm{accuracy} =
    \frac{TP_{int}}
    {(TP_{int} + FP_{int} + FN_{int})}
\end{equation}
where, 
\begin{equation*} 
    \rm{TP}_{int} =
    \frac{\rm{sum \: of \: true \: peak \: intensities}}
    {\rm{sum \: of \: predicted \: peak \: intensities}} 
\end{equation*}
\begin{equation*} 
    \rm{FP}_{Int} = 
    \frac{\rm{sum \: of \: false \: positive \: peak \: intensity}}
    {\rm{sum \: of \: predicted \: peak \: intensity}}  
\end{equation*}
\begin{equation*} 
    \rm{FN}_{int} =
    \frac{\rm{sum \: of \: false \: negative \: peak \: intensity}}
    {\rm{sum \: of \: ground \: truth \: peak \: intensity}}
\end{equation*}
The $\rm{TP}_{int}$, $\rm{FP}_{int}$, $\rm{FN}_{int}$ denotes intensity-weighted true positive peaks, false positive peaks and false negative peaks detected, respectively, from the predicted structure factor images. we note that the CBED and the structure factor images in our training dataset was generated with a pixel size of 0.0217 \AA{}$^{-1}$. To measure the intensity-weighted accuracy and the three metrics - $\rm{TP}_{Int}$, $\rm{FP}_{Int}$, $\rm{FN}_{Int}$ for predicted structure factor, we use a threshold size of 0.05 \AA{}$^{-1}$ to match peaks between the predicted and ground truth structure factor images, in order of peak pair distance. Several example diffraction images, sampled randomly from the test dataset, are shown in Fig.~\ref{Fig:disk_compare}a. The corresponding computed and ground truth disk positions and amplitudes are shown in Figs~\ref{Fig:disk_compare}b and c, using cross correlation and our trained FCU-Net, respectively. The accuracy of disk detection using the FCU-Net is significantly better than the correlation0based approach across the board, with the most striking gains occurring in diffraction patterns which suffer from multiple scattering due to large thickness, or disk overlap when the scattering vectors are small compared to the probe semiangle. 

The leftmost diffraction pattern in Figs~\ref{Fig:disk_compare}a is comparatively simple, with well separated, flat disks and signal well about the background level.  Unsurprisingly, both methods do very well.  However even here, in this nearly optimal data for cross correlative template matching, the gains using FCU-Net are remarkable, achieving 100\% accuracy.  In the middle three patterns, the background signal and disk overlap make visual identification of the disk positions difficult.  It is thus again unsurprising that cross correlation does relatively poorly.  In contrast, FCU-Net is extremely accurate for these three cases.  The fifth diffraction image in Fig.~\ref{Fig:disk_compare}a is an example of an experiment where the sample which has been tilted away from the low-index zone axis relative to the beam direction, creating complex variation in disk intensities due to tilt of the Ewald sphere. FCU-Net still outperforms cross correlation in this case, though the gains here are more modest.

We ascribe the improved accuracy of the FCU-Net to both the Fourier space convolutional layers which allow information from all lattice vectors to be pooled together, and to the large size of our training dataset.  Together, these enable the FCU-Net to correctly estimate the position of structure factor peaks even when the Bragg disks overlap, when signal-to-noise is low, or in the presence of nonlinear variation of the signal within the disks. We believe this robustness makes FCU-Net a good candidate for measurements of samples with unknown structures and orientations, where it may not be possible to guarantee non-overlapping disks or thick samples.

\begin{figure*}[htbp]
\includegraphics[width=6.0in]{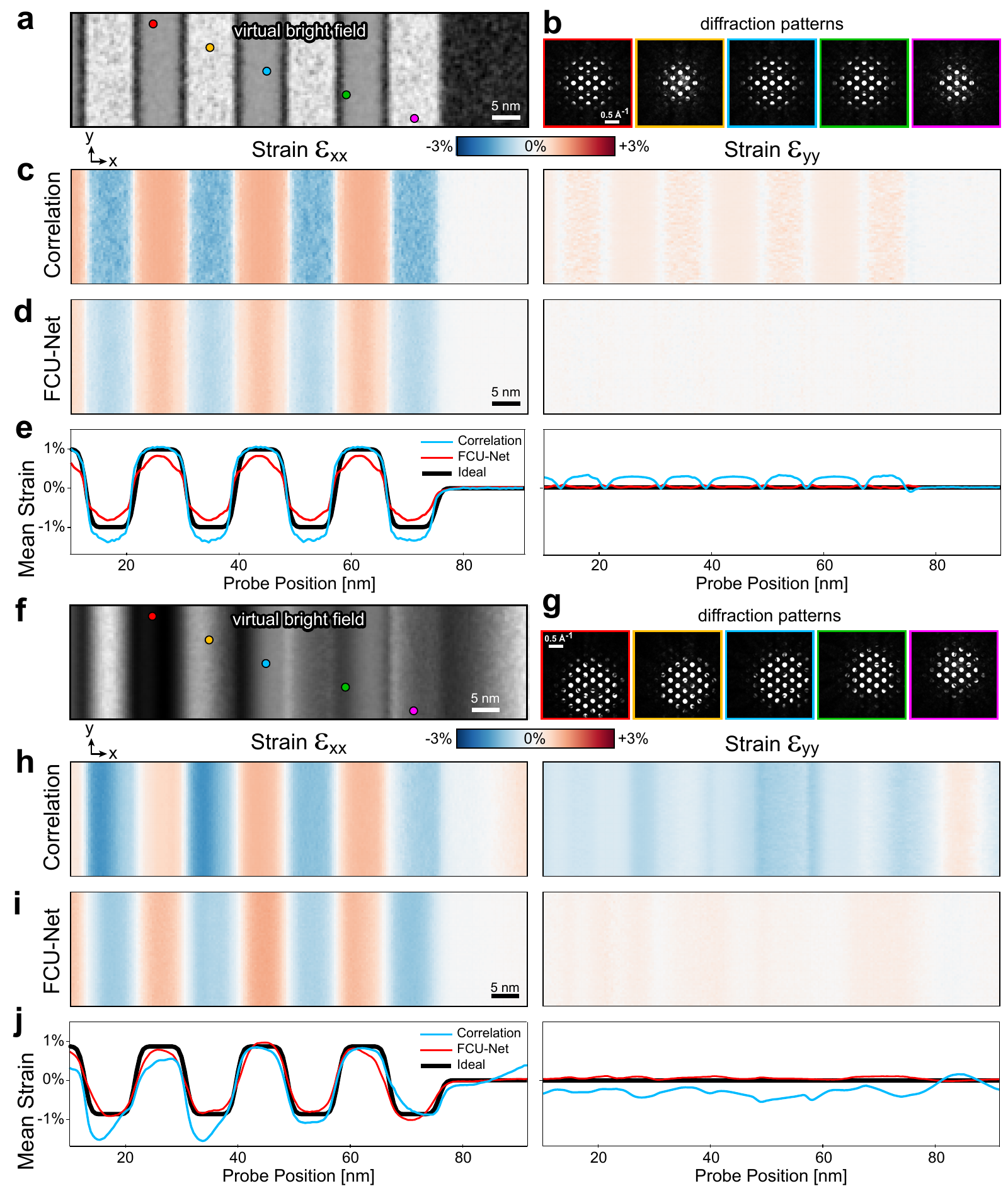}
\caption{{\bf Strain measurements from diffraction simulations of a Si-Si$_{0.5}$Ge$_{0.5}$ multilayer stack, (a)-(e) without mistilt, and (f)-(j) with helical mistilt.}
(a)/(f) Virtual bright field images calculated from the center disk, with the diffraction patterns corresponding to marked probe positions given in (b)/(g). Strain maps measured with (c)/(h) cross correlation and (d)/(i) FCU-Net. (e)/(j) Line profiles of the mean strain perpendicular (left) and parallel (right) to the interfaces.}
\centering
\label{Fig:simulated_strain}
\end{figure*}


\subsection*{Strain maps from simulated Si-SiGe multilayer data}

We next compare strain maps generated using both the cross correlation and FCU-Net approaches for realistic simulated datasets. The sample geometry consists of alternating layers of Si and SiGe on a mixed SiGe substrate. Two datasets are shown in Fig.~\ref{Fig:simulated_strain}, both containing the same strain profile, which alternates between $\pm 1\%$ strain relative to the substrate. The first, shown in Fig.~\ref{Fig:simulated_strain}a-e, is perfectly aligned along the [011] zone axis. The second, shown in Fig.~\ref{Fig:simulated_strain}f-j, has been helically twisted such that all regions of the sample are tilted away from the ideal diffraction condition. The tilt magnitude varies linearly from $0.4^\circ$ to $4.4^\circ$ from the substrate to the left side, and the tilt direction varies linearly from $45^\circ$ to $315^\circ$ relative to the x axis.

Fig.~\ref{Fig:simulated_strain}a shows a virtual bright field image constructed from the center disk across all the diffraction patterns in the perfectly aligned sample. Diffraction patterns from the five regions marked in Fig.~\ref{Fig:simulated_strain}a are presented in Fig.~\ref{Fig:simulated_strain}b. The strain maps for this sample along the two principal directions, $\epsilon_{xx}$ and $\epsilon_{yy}$, are plotted in Fig.~\ref{Fig:simulated_strain}c and d using the correlation method and the FCU-Net model, respectively. For both predictions, the reference lattice is set to be the mean lattice measured from the substrate region on the right hand side. 

Fig.~\ref{Fig:simulated_strain}e plots line profiles along the $x$-direction, perpendicular to the interfaces, of the mean strain for each of $\epsilon_{xx}$ and $\epsilon_{yy}$ (left and right, respectively). The strain parallel to the layer interfaces should be $\epsilon_{\rm{yy}}=0$ everywhere (for an epitaxial film). The $\epsilon_{\rm{yy}}$ strain estimated from correlation shows significant deviation from the expected zero strain value, varying systematically and periodically from zero strain near the interfaces, producing a RMS error of $\sim$ 0.2\% across the multilayer stacks. In contrast, the FCU-Net $\epsilon_{\rm{yy}}$ strain shows almost negligible systematic and random errors (RMS error $\leq$ 0.02\%). The strain in the normal direction $\epsilon_{\rm{xx}}$ should optimally follow the ideal profile plotted in Fig.~\ref{Fig:simulated_strain}e. Both approaches perform reasonably well, with the correlation method performing slightly better in the positively strain layers (tension) while the FCU-Net underestimates the strain magnitudes at the middle of each layer.


Interestingly, in the perfectly on-zone crystal, FCU-Net systematically underestimates $\epsilon_{xx}$ within each layer, and rounds off the sharp interfaces between layers. Importantly, this effect was not present in the simulated distorted sample or the experimental data sets, as will be discussed in subsequent sections. We surmise two possible sources of this error. The first is the complexity of electron scattering along high symmetry zone axes. In on-zone scattering, effects like electron channeling and increased beam coherence make it inherently difficult to generalize or predict \textit{a priori} how the internal structure of Bragg disks will vary from one sample to the next, or even between two samples which are identical save for a single additional atomic layer. Half of the training data was aligned along a low-index zone axis, and we hypothesize that 100,000 unique diffraction patterns may be sufficient to train FCU-Net well for off-axis electron scattering, but be insufficient for on-axis training. The second possible source of the error is the presence of an interface, which will affect the diffraction patterns even more strongly on-zone. The training data contained only pure crystals, therefore fine tuning the FCU-Net model with complex geometries such as diffraction at an interface would improve its accuracy in predicting strain maps for samples with interfaces.


Next, we calculate strain maps from the simulated multilayer dataset which has been twisted off the ideal diffraction condition. Fig.~\ref{Fig:simulated_strain}f shows the virtual bright field image, and Fig.~\ref{Fig:simulated_strain}g plots the diffraction patterns for selected positions marked in Fig.~\ref{Fig:simulated_strain}f.  The varying stripes of intensity in the bright field image, and the shifting disk intensity envelope function in the five shown diffraction patterns, both result from the helical twisting of the sample.  We again calculate strain maps along the principal directions, shown in Fig.~\ref{Fig:simulated_strain}h (correlation) and i (FCU-Net). Once again, the reference lattice for the calculation was taken to be the mean lattice vectors from the substrate region on the right of the scan.

 Fig.~\ref{Fig:simulated_strain}j plots the line profile of mean strain values parallel and perpendicular to the multilayer stacks. The expected strains are again $\epsilon_{\rm{yy}}=0$, and $\epsilon_{\rm{xx}}=\pm1\%$ alternating between the Si and SiGe layers.  In $\epsilon_{yy}$, the estimates from the correlation method deviate significantly from 0 strain, with a RMS error of approximately $0.6\%$ in the multilayer region. By contrast, the FCU-Net predictions are closer to the expected zero strain value, with a negligibly small RMS error ($<$ 0.1\%).  In $\epsilon_{\rm{xx}}$, the correlation method is accurate for several of the layers close to the middle of the scan region, where the mistilt is smallest; however, it becomes quite inaccurate on the left half of the image, where it captures the location of the interfaces but systematically and significantly underestimates the true strain values and fabricates variation within individual layers, where the profile should be flat.  Similarly, correlation becomes inaccurate on the far right of the image, in the reference substrate, making it challenging to even estimate the reference lattice.   We attribute these artifacts to the varying tilt of the sample, which is known to deleteriously affect template matching by shifting the center of mass of disk intensities.  In contrast, the FCU-Net $\epsilon_{xx}$ strain map mirrors the ground truth value with good fidelity, showing only small deviations such as some slight rounding of the interfaces.  The effectiveness of FCU-Net in the presence of sample mistilts is important, as this is a common occurrence in experimental data and very often creates significant artifacts using traditional methods.


\subsection*{Strain maps from experimental h-BN films}

\begin{figure}[htbp]
    \centering
    \includegraphics[width=3.4in]{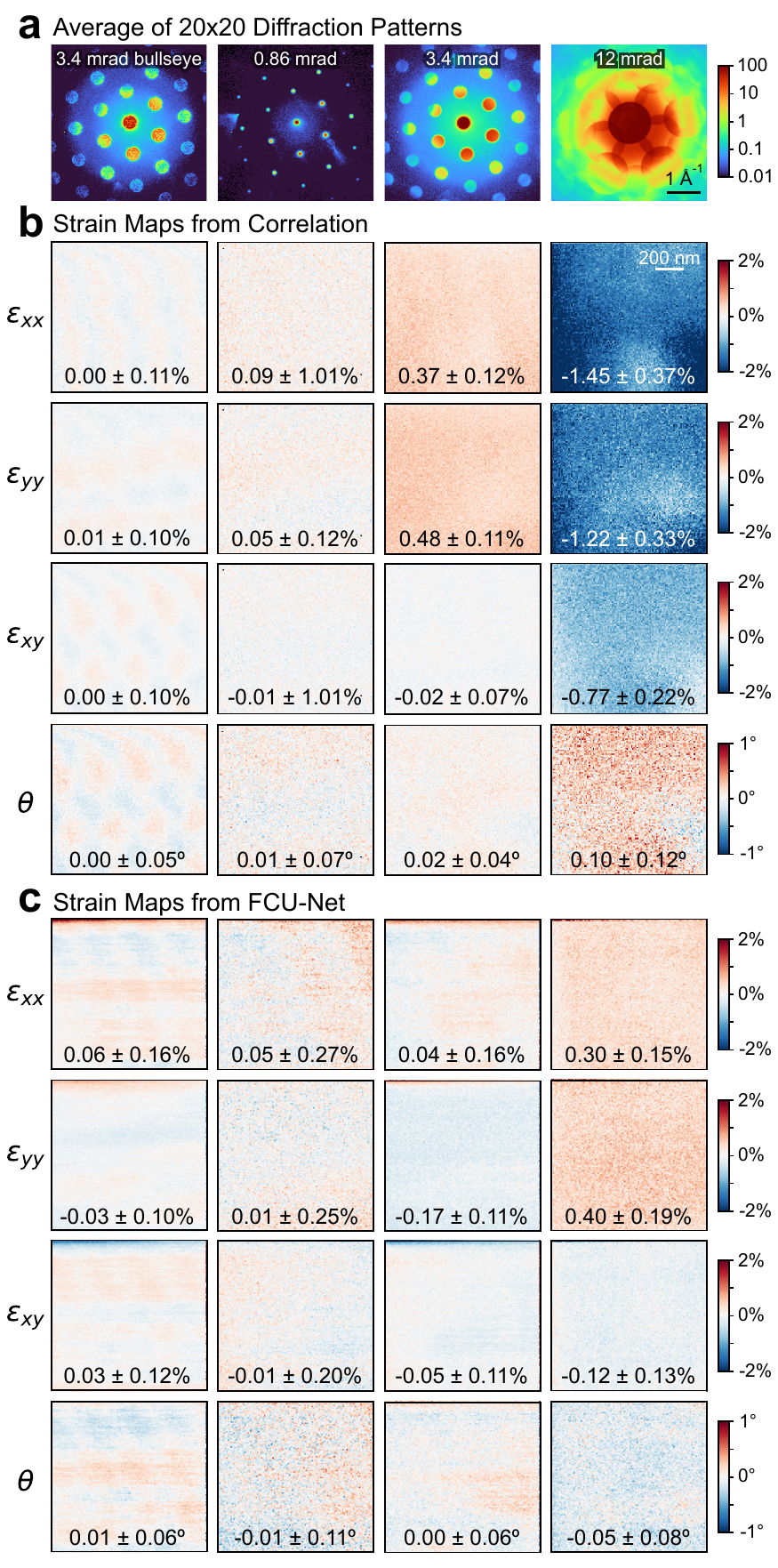}
    \caption{{\bf Experimental strain maps measured from single crystal hexagonal-boronitride thin films.} (a) Mean diffraction images of 20x20 probe positions, for STEM probes defined by 3.4, 0.86, 3.4 and 12 mrad semiangle apertures, where the leftmost aperture also contains a bullseye pattern. (b) Strain maps measured using cross correlation template matching for the 4 cases given above. (c) Strain maps measured using the FCU-Net network predictions. For all maps, the mean and standard deviation strains/angles are inset.}
    \label{Fig:StrainMap_hBN}
\end{figure}

\begin{figure*}[htbp]
    \centering
    \includegraphics[width=6in]{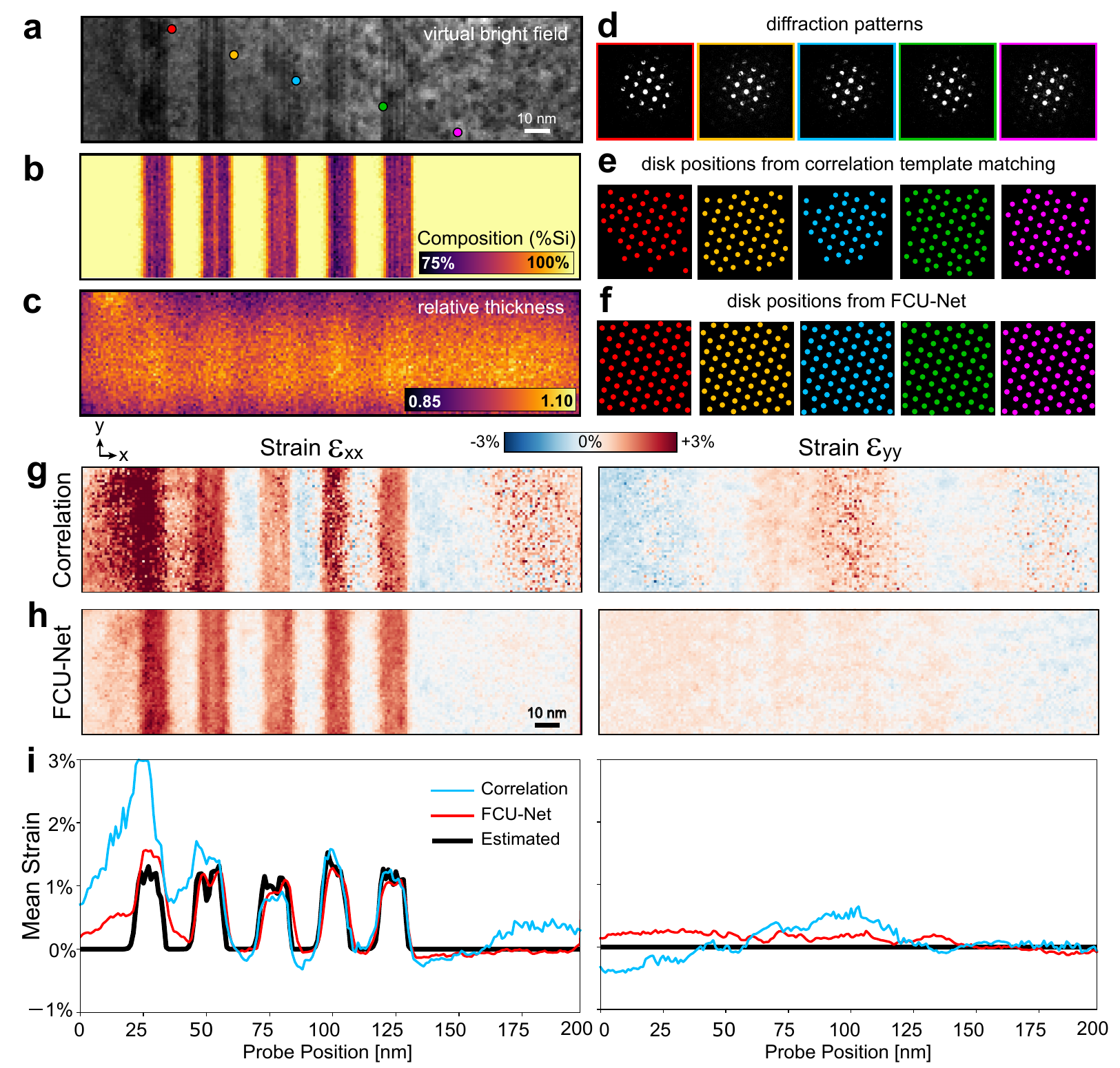}
    \caption{{\bf Experimental strain and composition characterization of a Si-Si$_{0.87}$Ge$_{0.13}$ multilayer stack.} (a) Virtual bright field calculated from center disk. (b) Composition and (c) relative thickness, estimated from STEM-EELS. (d) Diffraction patterns corresponding to the probe positions marked in (a), with estimated Bragg disk positions from (e) correlation template matching and (f) FCU-Net. (g) Strain maps measured from correlation template matching. (h) Strain maps measured from FCU-Net. (i) Mean strain values parallel to the multilayer normal direction, for correlation, FCU-Net, and estimated from the STEM-EELS composition.}
    \label{Fig:StrainMap_exp}
\end{figure*}

To test the performance of FCU-Net on experimental data, We compute strain maps for experimental hexagonal boron nitride (h-BN) data using cross correlation and FCU-Net.  Data was collected using four different electron probes, three with circular apertures and convergence semiangles of 0.86, 3.4 and 12 mrads, and one with a bullseye patterned aperture and 3.4 mrad semiangle \citep{zeltmann2020patterned}.  Fig.~\ref{Fig:StrainMap_hBN}a shows mean diffraction patterns from 20$\times$20 different scan positions for each of these probes.  Figs.~\ref{Fig:StrainMap_hBN}b and c show strain maps from the correlation and FCU-Net methods, respectively, with the reference lattice set to the average of all positions in the bullseye pattern measurements. The full strain tensor is shown for all positions, consisting of the two principal strain direction $\epsilon_{\rm{xx}}$ and $\epsilon_{\rm{yy}}$, the shear strain $\epsilon_{\rm{xy}}$, and the rotation $\theta$. We expect the single crystal h-BN sample to be essentially free of strain and local rotations, suggesting an ideal measurement of 0 for all channels. The mean and standard deviation of the strain values for all probe positions (excluding the first two scan rows) are inset into each panel in Figs.~\ref{Fig:StrainMap_hBN}b and c. The mean and standard deviations represent the systematic and random errors respectively. Because the field of view is so large, there is some thickness and tilt variation over the field of view.

The first column of Figs.~\ref{Fig:StrainMap_hBN}b and c shows results from the 3.4 mrad bullseye probes.  Cross correlation and FCU-Net both perform very well on this data, producing means and standard deviations very close to zero.  Some position dependent systematic errors are visible for both methods, possibly due to the sharp edges of the patterned aperture combined with the few pixel shifts of the patterns over the field of view. Interestingly, it is worth noting that FCU-Net does quite well with the bullseye data, despite being trained only on normal, circular probes. The surprisingly impressive performance in the strain measurements with completely unseen diffraction images from patterned aperture can be attributed to the introduction of the Fourier space cross correlation preprocessing layer as implemented in the FCU-Net model (Fig.\ref{Fig:FCUnet}). 

Similarly, for the 0.86 mrad probes, shown in the second column of Figs.~\ref{Fig:StrainMap_hBN}b and c, both correlation and the FCU-Net perform well overall, with means close to 0 in all cases.  The standard deviations, indicating the random error, are larger than for the bullseye data, with values as high as $\sim 1\%$ for the correlation $\epsilon_{xx}$ and $\epsilon_{xy}$ maps and ~ $0.25\%$ for several of the FCU-Net maps.

These first two columns represent experimental conditions that are well suited to Bragg disk detection using cross correlation. Bullseye apertures were specifically designed to perform disk detection well using template matching, and this result is borne out here; however, these apertures sacrifice spatial resolution and introduce high frequency components to the probe shape in real space. Similarly, using a small convergence semi-angle improves the disk detection accuracy with cross correlation by minimizing the chance of disk overlap and the effects of intensity variation within the disks, at the cost of limiting the spatial resolution since reducing the probe size in diffraction space increases its size in real space. The capacity to accurately detect disk position while opening up the aperture size is therefore highly desirable if high spatial resolution is required.




In the third column of Figs.~\ref{Fig:StrainMap_hBN}b and c (3.4 mrad probe), the disks begin to show significant intensity gradients within the disks, with higher intensities closer to the origin. This leads to significant positive systematic error in the principal strains ($\epsilon_{\rm{xx}}$ and $\epsilon_{\rm{yy}}$) for the correlation estimates. This is likely because the correlation-estimated disk positions are slightly biased towards the origin, leading to a smaller estimated reciprocal lattice and thus positive real space strains. This effect should not modify the results for either shear strain or rotation, and indeed both of these quantities show low error. By contrast, the FCU-Net predictions show low systematic errors for all 4 components of the strain tensor, demonstrating the robustness of the FCU-Net approach to variations in disk intensities. Both methods show fairly low random errors of $0.10\%$ and $0.13\%$ for correlation and FCU-Net respectively.

In the final column of Figs.~\ref{Fig:StrainMap_hBN}b and c (12 mrad probe), the disks have expanded to create significant overlap, a condition required for atomic resolution imaging, but which typically thwarts traditional template matching.  The resulting systematic errors are very high, approximately $-1.1\%$, and significant variation over the field of view is visible in all correlation measurements. FCU-Net, in spite of being trained on images with probe semiangles up to a maximum of 4 mrads, performs fairly well on this data, with systematic errors approximately 3 times lower than the correlation method. We ascribe this to the training dataset containing many crystals and orientations that produce disk overlaps for 4 mrad probes (and below), such that the network has learned to interpret the nonlinear interference patterns formed in the presence of overlapping disks. The random errors are also lower for the FCU-Net compared to the correlation method, and the predicted strains show less variation across the field of view. Overall, the FCU-Net produces more accurate and precise strain predictions over a wider parameter range than the correlation method, including experimental conditions it was not exposed to during training. We also note that the strain measurement accuracy using FCU-Net model may be further improved by fine tuning the pre-trained model with application specific diffraction data.  

\subsection*{Strain maps from experimental SiGe multilayer stacks}

Finally, we compare the two strain calculation methods on a thick, non-uniform multilayer stack of alternating layers of Si and a mixture of Si and Ge grown epitaxially. A virtual image constructed from the center disk is shown in Fig.~\ref{Fig:StrainMap_exp}a. We observe significant contrast differences over the field of view, corresponding to variation in the sample's thickness, composition and surface morphology. We have estimated the local composition of the sample by using STEM-EELS, shown in Fig.~\ref{Fig:StrainMap_exp}b. The mean composition of the 5 stripes from STEM-EELS is Si$_{0.82}$Ge$_{0.18}$. We estimate that the average thickness of the sample is $\approx$ 110 nm, using the $t/\lambda$ method \cite{malis1988eels} applied to the pure Si regions and are therefore in the multiple scattering regime \cite{vallejo2018observation}. The local relative thickness is plotted in Fig.~\ref{Fig:StrainMap_exp}c, showing a relative thickness variation of about 20\%.

We plot examples of the diffraction patterns in Fig.~\ref{Fig:StrainMap_exp}d, from 5 regions marked in Fig.~\ref{Fig:StrainMap_exp}a. We see significant variation in the fine structure of the diffracted disks, especially when comparing regions of different compositions. The round shape of many of the disks are significantly degraded due to the thickness and non-uniformity of the sample. Finally, the center-of-mass of the diffraction pattern intensities changes over the field of view, indicating that bending of the sample had lead to slightly different tilt conditions for different probe positions. We have used both cross correlation and FCU-Net to estimate the Bragg disk positions, with examples shown in Figs.~\ref{Fig:StrainMap_exp}e and f, corresponding to the diffraction patterns shown in  Fig.~\ref{Fig:StrainMap_exp}d. The resulting disk positions are noticeably less regular for the correlation method, and many disks at higher diffraction angles close to the image edges are too weak to be identified. This is in contrast to the FCU-Net predictions, which returns a highly regular lattice of disk positions, with only a few weak false positives visible at the image boundaries.

The strain maps along the principal directions calculated with the correlation method are shown in Fig.~\ref{Fig:StrainMap_exp}g, and those calculated using the FCU-Net predictions are shown in Fig.~\ref{Fig:StrainMap_exp}h. In both cases, the reference lattice was taken to be the mean lattice vectors from the substrate region on the right of the field of view. Fig.~\ref{Fig:StrainMap_exp}i plots line profiles of the mean strain values perpendicular (left) and parallel (right) to the multilayers. In the parallel direction, we expect the strain will be $\epsilon_{\rm{yy}}=0$ everywhere, due to the epitaxial nature of the layers. The correlation strain shows significant deviation from 0 strain, and moreover, is not flat over the imaged area, with deviations ranging from approximately -0.4\% on the left side, to +0.6\% in the center, and back down to 0\% in the substrate region on the right hand side. The FCU-Net strain $\epsilon_{\rm{yy}}$ by contrast is comparatively flat, and ranges from approximately +0.2\% on the left side, to 0\% strain in the substrate on the right hand side. We note that while the RMS error in strain $\epsilon_{\rm{yy}}$ calculation across all the multilayer stacks is $\sim$ 0.3\% with cross correlation approach, it is $\sim$ 0.15\% from the FCU-Net prediction.

In the normal direction, we can compare the strain $\epsilon_{\rm{xx}}$ computed with cross correlation and with FCU-Net to the strain measured using independent STEM-EELS measurements.  The STEM-EELS result is shown as a black line in Fig.~\ref{Fig:StrainMap_exp}i.  The FCU-Net line profile closely approximates the STEM-EELS profile, capturing most of the sharp transitions at the interfaces, and the roughly flat profiles within each layer.  The cross correlation result fares much worse, capturing the $\epsilon_{xx}$ structure of the three right-most layers roughly correctly, but then deviating wildly on the left side of the scan region, possibly due to local sample mistilt.  The correlation result also deviates from a flat profile in the substrate on the right, making identification of a reference lattice difficult. For the strain $\epsilon_{\rm{xx}}$, FCU-Net produces a RMS error of approximately 0.25\% across the sample leading to almost three-fold increase in the accuracy from cross correlation, which produced a RMS error of approximately 0.72\%. This example highlights common pitfalls of traditional template matching in the presence of complex, nonlinear electron scattering signals, and the capacity of the FCU-Net model to achieve accurate disk localization measurement in spite of these challenges.



In summary, we have developed a deep learning network  (FCU-Net) for quantitative measurements of Bragg disk positions from electron diffraction patterns. Our networks have been trained with over 200,000 unique, simulated diffraction patterns with thicknesses ranging from 2 to 50 nm thick, covering more than 1000 distinct crystal systems over many orientations and microscope parameters. We found that the resulting Bragg disk position predictions from the FCU-Net network were substantially more accurate than a conventional template matching correlation method. We tested the FCU-Net predictions for crystalline lattice strain mapping, using both simulated and experimental 4D-STEM datasets. In both cases, we found that the FCU-Net predictions were substantially more robust against signal variations due to mistilt of the sample and multiple scattering due to sample thickness. We have integrated FCU-Net into the open source 4D-STEM analysis python library \pyFDSTEM{}, providing free access and use of the network, and a complementary suite of tools for subsequent analysis of the measured structure factors, to the electron microscopy community. All of our simulated and experimental datasets, source codes, and trained networks are freely available in open source repositories. The improved accuracy and precision of Bragg disk measurements using FCU-Net, even in the presence of complex signals involving thick samples and multiply scattered electrons, can provide widespread benefits in 4D-STEM application such as strain, phase, and orientation mapping, and in quantitative electron crystallography.


\section*{Methods}
\label{section:methods}

Fig.~\ref{Fig:overview} shows a flow chart of the methods we use to invert STEM diffraction patterns into quantitative structure factor positions and amplitudes. First we generate a library of simulated dynamical diffraction data  (Fig.~\ref{Fig:overview}a). We selected thousands of unique material systems that span a wide variety of crystallographic prototype systems, and simulated the CBED patterns at various thicknesses, tilts, and microscope conditions using the multislice algorithm  \cite{cowley1957scattering, kirkland2020advanced}. The projected structure factors are then computed, including the effect of any excitation error by evaluating the distance of the projected potentials from the Ewald sphere.  Simulated data which will be used for training is then augmented with noise profiles which mimic real experimental conditions.  The network is then trained using the noise-augmented simulated data.  Fig.~\ref{Fig:overview}c overviews the input, architecture, and output of the FCU-Net deep neural network used to predict the (projected) structure factor positions from the input diffraction patterns and electron probe.
Figs.~\ref{Fig:overview}d-f show the typical inference stage, where we use the pre-trained FCU-Net model to predict the underlying structure factor positions and amplitudes from experimental diffraction patterns.


\subsection*{Dynamical diffraction library simulations}

\begin{figure*}[htbp]
    \centering
    \includegraphics[width=1.04\textwidth]{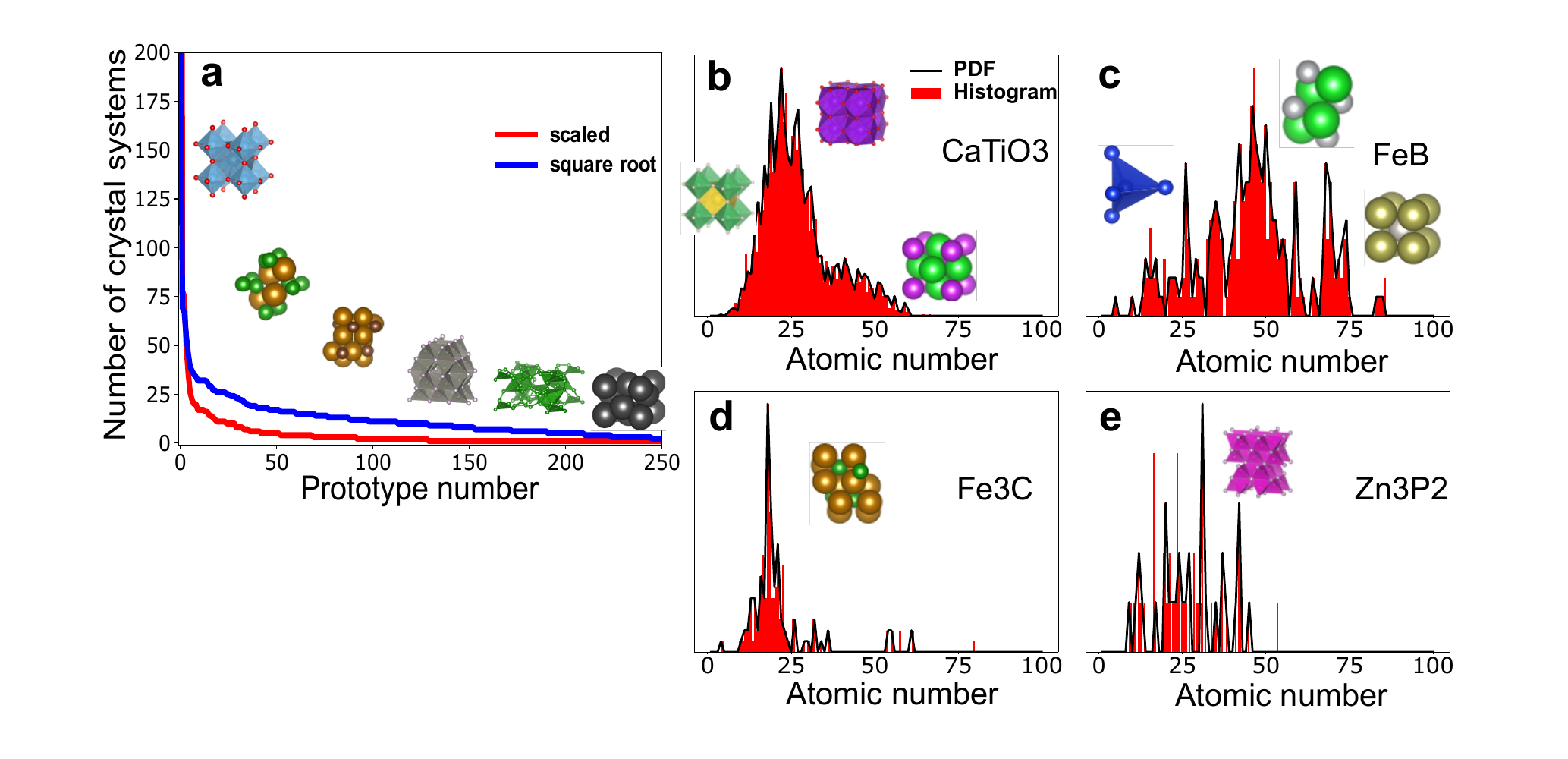}
    \caption{{\bf Crystal system extraction from the materials project database.} (a) Number of crystal systems chosen from each prototype systems for the training dataset. (b)-(e) Atomic number distribution of crystal systems belonging to the same prototype system as (b) CaTiO$_3$, (c) FeB, (d) Fe$_3$C, (e) Zn$_3$P$_2$.}
    \label{Fig:CrysAflow}
\end{figure*}

To build a dynamical diffraction library for the AI/ML training, we implemented an automated pipeline which selects the crystal structures, and simulates CBED patterns and the underlying projected structure factors with a variety of experimental parameters. The dynamical diffraction library generation starts with building a materials database. To judiciously select crystal structures of interest for our problem, we initially compare $\approx 139,000$ crystal structures and compositions from the materials project (MP) database \cite{Jain2013} with more than 500 crystallographic prototypes collected from the AFlow library (Fig.~\ref{Fig:CrysAflow}) \cite{mehl2017aflow,hicks2019aflow}. Crystallographic prototypes are an alternative and popular crystal structure classification paradigm.
Fig.~\ref{Fig:CrysAflow}a shows the distribution of the crystal systems from the MP database, grouped according to their structural similarity with crystallographic prototype systems. We presented the first 250 prototype systems, as shown in Fig.~\ref{Fig:CrysAflow}a, which cumulatively span approximately 95\% of the materials systems from materials project database. We sampled $\sim$ 1000 unique crystal systems following the distribution, presented as a blue line in Fig.~\ref{Fig:CrysAflow}a. 

Figs.~\ref{Fig:CrysAflow}b-e plots the distribution of atomic number space of the crystal structures which are structurally similar to four different example prototype systems - CaTiO$_3$, FeB, Fe$_3$C and Zn$_3$P$_2$. As evident from the distribution in panel b-e, the selected materials systems have diverse range of constituent atomic elements. Following the crystal system extraction, we simulated the CBED patterns and underlying structure factors using the multislice algorithm \cite{cowley1957scattering, kirkland2020advanced}, as implemented in the \prismatic{}  code \cite{ophus2017fast, dacosta2021prismatic}.

From these simulations, the corresponding ground truth structure factors are calculated from the projected atomic potentials for each diffraction pattern. This is achieved by first transforming atomic potentials into 3D Fourier space, applying a 2D Tukey window function in the projection plane, and 2D Fourier downsampling to attain the desired output resolution in x and y. A Gaussian weighted filter is applied along z axis (the beam direction) with a standard deviation of $0.05$ \AA{}$^{-1}$ to select the structure factors close to the projection slice. Finally, the projection is summed along z axis to generate the ground truth structure factors. Note that these structure factor images are depend linearly on the thickness of the sample. We simulated CBED patterns and the underlying structure factors for all the 1000 unique crystal systems for thicknesses between 2 to 50 nm with an interval of 2 nm. For each crystal system we simulated diffraction patterns for the crystal orientated along 5 different low-index zone axes, and 5 random orientations. We simulated diffraction patterns for each orientation with probe semiangles of 1, 2, and 4 mrads. In total this yielded diffraction library of 750,000 diffraction patterns, each with a unique combination of crystal system, sample tilt, specimen thickness and probe convergence angle. For each of the 750,000 diffraction patterns the probe and structure factors were also created. We have implemented a parallelized framework for the data simulation, training data generation, and training steps \cite{rakoski2021database}.

\subsection*{Conventional Bragg disk position measurements}

Determining the Bragg disk positions and intensities in each diffraction pattern is an important step which allows subsequent measurement of parameters such as phase, orientation, and strain in crystalline and semi-crystalline materials. Cross-correlative template matching is one method routinely used to measure the positions of Bragg disks \citep{ozdol2015strain, pekin2017optimizing}, matching to either raw diffraction patterns or edge-filtered images \cite{mukherjee2020lattice}. In the template matching approach, the Bragg disk positions are calculated in two steps - first, we collect the undiffracted probe over vacuum to create our template for matching. Next we perform cross correlation between the diffraction pattern and the probe template in Fourier space  to find all disk positions in a given diffraction pattern. In this work, we use the disk detection, lattice fitting, and strain mapping tools implemented in the open source python package \pyFDSTEM{} \cite{savitzky2021py4dstem}.


\begin{figure*}[htbp]
\includegraphics[width=6.0in]{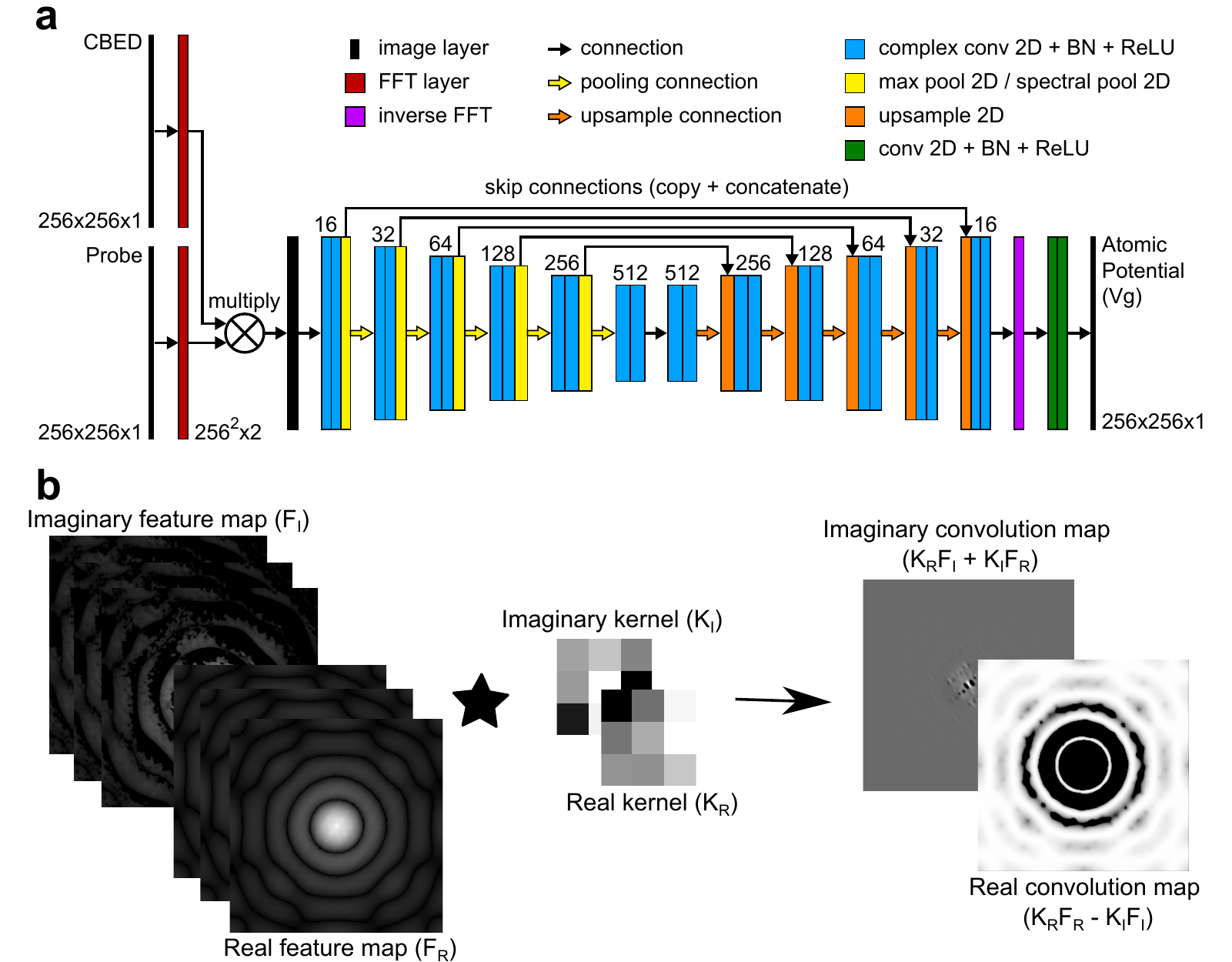}
\caption{{\bf FCU-Net network architecture.} (a) Architecture of the neural network implemented to predict pixel-wise regression maps of the projected atomic potential. (b) Complex convolution operation performed on CBED images cross-correlated with vacuum probe template.}
\label{Fig:FCUnet}
\centering
\end{figure*}


\subsection*{Bragg disk detection using Fourier space deep learning}

We implement three variants of CNN architecture - U-Net \cite{ronneberger2015unet}, and its modified variants with spectral parameterization adapted from Ripple \textit{et al.} \citep{rippel2015spectral} and fully complex variant, FCU-Net adapted from Trabelsi \textit{et al.} \citep{trabelsi2017deep}. Fig.~\ref{Fig:FCUnet}a presents the model architecture of U-Net and its hybrid variants with fully complex convolution and spectral pooling layers. The FCU-Net architecture implemented in this work considers two inputs: the probe template and the CBED diffraction pattern. To make the FCU-Net model aware of the vacuum probe template, we implement a preprocessing layer which multiplies the Fourier transform of the diffraction pattern with the probe template. Finally, we implement the 2D complex convolutional layer, which is the building blocks for the FCU-Net, to teach the complex space information from the Fourier transformed image from the pre-processing layer. Following a combination of complex convolutions, pooling and upsampling operations the final output from the FCU-Net is transformed using inverse Fourier transform operation, before it is compared with the ground truth atomic potentials. 

\subsection*{Complex convolution}

We implement complex convolutional layers by independently initializing real and imaginary components of the 2D convolutional kernel (Fig.~\ref{Fig:FCUnet}b), that is, we consider the real and imaginary parts of the complex numbers as logically distinct real-valued numbers. Akin to the 2D real-valued convolution operator, we convolve a complex kernel matrix (\(K = K_R + iK_I\)); $K_R$, $K_I$ $\in$ $\mathbb{R}^{m/2\times m/2}$  with the complex input feature map (\(F = F_R + iF_I\)); $F_R$, $F_I$ $\in$ $\mathbb{R}^{m/2\times N}$, where $m/2$ is the size of the complex kernel weight and $N$ is the number of pixels in the input image (feature map). The complex convolution operation can be formulated as:
\begin{equation}
    K * F 
    = 
    \left(
        K_R*F_R - K_I*F_I
    \right)
    +i(K_I*F_R + K_R*F_I),
\end{equation}
We can use a matrix notation to represent the complex convolution operator: 
\begin{equation}
\begin{bmatrix}
\mathfrak{Re}(K*F)\\
\mathfrak{Im}(K*F)
\end{bmatrix}
= 
        \begin{bmatrix}
K_R & -K_I\\
K_I & K_R
\end{bmatrix}
    *\begin{bmatrix}
F_R\\
F_I
\end{bmatrix},
\end{equation}
Out of the variety of options available for activation functions for complex convolutions, we have chosen to use the complex rectified linear unit  ($\mathbb{C}$ReLU) function such that for any complex number $z$ :
\begin{equation}
    \mathbb{C}ReLU(z)
    = 
    ReLU(\mathfrak{Re}(z)) +
    i ReLU(\mathfrak{Im}(z)),
\end{equation}
Trabelsi \textit {et al.} recently compared different variants of ReLU functions for complex operators, and found that $\mathbb{C}$ReLU(z) had the best performance \cite{trabelsi2017deep}. In our tests, we found $\mathbb{C}ReLU(z)$ to be the preferred nonlinear activation function, as it can distinguish correlations from the complex convolution operation into four distinct region based on if the $\mathfrak{Re}(z)$ and $\mathfrak{Im}(z)$ are strictly positive or negative. For deep networks such as FCU-Net, this provides the required flexibility and nonlinearity to the network by allowing complete manipulation of the phase information at each layer of the network.

\subsection*{Spectral pooling}

To implement the U-Net with spectral parameterization we replace the max-pooling layers typically used in U-Nets with spectral pooling layers as we find that this reduces the introduction of artifacts and nonlinearity, resulting in a more stable and accurate prediction from the network. Where max-pooling layers down sample the image in real space, spectral pooling operates in the frequency domain. Spectral pooling in its original form as described by Rippel et. al., \cite{rippel2015spectral} transforms an image to Fourier space by applying a fast Fourier transform operation (FFT), after which it is cropped in Fourier space and transformed back to real space by an inverse FFT such as:
$x \in \mathbb{C/R}^{M \times {M}}  \xrightarrow[]{\text{FFT}} \tilde{x} \in \mathbb{C}^{M \times {M}}  \xrightarrow[]{\text{Crop}} \tilde{x} \in \mathbb{C}^{N \times {N}}  \xrightarrow[]{\text{inv FFT}} x \in \mathbb{C}^{N \times {N}}$, where $x$ and $\tilde{x}$ are the input and Fourier transformed image respectively, N and M correspond the number of pixels in the image, with $N < M$.


\subsection*{Training FCU-Net}


We train the fully complex FCU-Net network on the simulated sets of images composed of a vacuum probe, a CBED pattern, and the ground truth structure factors, for different material systems at different sample thicknesses up to 50 nm. To make FCU-Net robust against various experimental conditions, we augment the simulated images with several forms of noise typically found in 4D-STEM data: \textit{(i)} elliptical distortion and \textit{(ii)} random translations (x,y pixel shifts) of the diffraction patterns, \textit{(iii)} incoherent backgrounds modeled as plasmonic signal, \textit{(iv)} shot (counting) noise using Poisson statistics, and \textit{(v)} random bright (hot) and dark (dead) pixels to simulate the effect of X-rays and detector pixel errors.



For the final training, we randomly sampled $\sim$ 200,000 unique training ($\sim$ 20,000 test) triplets from the diffraction pattern library. Each triplet contained a vacuum probe and a CBED pattern, used as the training inputs and the structure factors for the training output. 
Table \ref{table:FCU-NetParameters} summarizes the hyperparameters considered during the FCU-Net training. Before the final training iteration, we implement a high-throughput hyperparameter optimization scheme using RayTune python library for deep learning \citep{liaw2018tune}. A random subset of the training data was used during hyperparameter tuning, as a compromise between accuracy and the computational overhead. Following the hyperparameter optimization, we perform the final round of training iterations for the FCU-Net on 8 NVIDIA Tesla V-100 (16 GB VRAM) GPU nodes using a distributed Tensorflow strategy to accelerate the training performance \citep{abadi2016tensorflow}. All training and test runs for this work were performed on the super-computing facility (Cori GPU clusters) at the National Energy Research Scientific Computing Center (NERSC). 

\begin{table}[h!]
\caption{Selected hyperparameters for FCU-Net deep neural network}
\begin{tabular}{m{4.7cm}| m{2.5cm}} 
 \hline
Hyperparameters &  \\ [0.5ex] 
 \hline
 Batch size & 256  \\ 
 Filter size & 32   \\
 Filter depth & 4  \\
 Drop out rate & 0.3  \\
 Activation & $\mathbb{C}ReLU$  \\
 \hline
\end{tabular}
\label{table:FCU-NetParameters}
\end{table}

\subsection*{Integration with py4DSTEM}

Bragg disk detection using the trained FCU-Net model is implemented in the \pyFDSTEM{} python data analysis toolkit developed by Savitzky \textit{et al.} \cite{savitzky2021py4dstem}. The workflow for AI/ML guided disk detection using py4DSTEM starts with loading a 4D dataset and the corresponding vacuum probe. These inputs are passed to a function which feeds them into the trained FCU-Net model, which returns the predicted disk positions.  Currently we host the latest (and previously archived versions) of pre-trained model weights on a cloud location and which is updated periodically with new weights with improved test performance. When called, the py4DSTEM AI/ML disk detection function will search for the latest FCU-Net weights and automatically download them prior to disk detection. Once the prediction is completed, we convert the predicted output (a 2D image-like array of structure factors) to a set of M peaks defined by the values $(q^x_m, q^y_m, I_m)$, which can be used with any of the existing downstream analysis modalities built into py4DSTEM.

\subsection*{Strain mapping}

Strain mapping was performed using py4DSTEM.  Using the measured disk positions, either from FCU-Net predictions or cross correlation, we fit the lattice vectors at each beam position.  A reference lattice is chosen, and the difference between the reference and local lattice vectors are then used to calculate the infinitesimal strain tensor
\begin{equation}
    \epsilon = \begin{pmatrix}
               \epsilon_{xx}  &  \epsilon_{xy} \\
               \epsilon_{yx}  &  \epsilon_{yy}
               \end{pmatrix}
\end{equation}
where $\epsilon_{xx}$ and $\epsilon_{yy}$ are the strain along the $x$ and $y$ directions, and $\epsilon_{xy}$ is the shear strain.  We additionally calculate $\theta$, the rotation of the local lattice relative to the reference lattice.  The selection of reference lattice is specified for each strain map computed.  More details can be found in \cite{pekin2017optimizing, savitzky2021py4dstem}.

\subsection*{Simulated diffraction of SiGe multilayers}

In order to test the robustness of our network for realistic samples, we perform simulations of thick samples which incorporate multiple scattering of the electron beam. The sample geometry we used is a multilayer stack along the [011] direction, composed of alternating Si and Si$_{0.5}$Ge$_{0.5}$ layers, on a Si$_{0.75}$Ge$_{0.25}$ substrate, where each phase has diamond cubic structure.
For ease of comparison of our measured strain values with the ground truth, we used slightly different lattice constants from known experimental values, setting the substrate to have a lattice parameter of 5.6034 \AA{}, and the multilayers to have precisely $\pm 1\%$ strains relative to the substrate.

\subsection*{Experimental diffraction of SiGe multilayers and h-BN films}

Experimental 4D-STEM datasets were acquired using the TEAM I instrument at the National Center for Electron Microscopy facility of the Molecular Foundry, a double aberration corrected Thermo Fisher Titan fitted with a Gatan Continuum energy filter and K3 direct electron detector. The K3 detector was operated in electron counting mode. Electron diffraction patterns were acquired in energy-filtered mode with a 15 eV slit centered on the elastic energy to suppress background noise from inelastic scattering.

\textit{Hexagonal-boron nitride:} In order to obtain a reference dataset from a thin, single crystal material with minimal characteristic strain we used thin a flake mechanically exfoliated from a single crystal of hexagonal boron nitride.  This flake was transferred to a silicon nitride TEM grid for 4D-STEM experiments. Multiple 4D-STEM datasets were acquired at an 80 kV accelerating voltage using four different apertures to compare algorithmic performance under various experimental conditions.  Three circular apertures were used, with convergence semiangles of 0.86, 3.4, and 12 mrad, and one bullseye-patterned aperture was used \citep{zeltmann2020patterned}, with a 3.4 mrad convergence semiangle.  For each aperture, data was acquired with a 50 ms dwell time, step size of 100~\AA{}, and scan size of 112$\times$108 probe positions. Diffraction patterns were binned 4x4 after electron counting. 


\textit{Si-Si/Ge multilayers:} In order to obtain an experimental dataset with a large and known strain, we used a silicon/silicon-germanium ``MAG$^*$I$^*$CAL'' calibration sample obtained from Ted Pella, Inc. The sample consists of a Si wafer with several layers of approximately 10~nm of Si/Ge mixture grown epitaxially. The sample is prepared for TEM as a polished cross-section with the [110] zone axis normal to the foil. Data was acquired at a 300~kV accelerating voltage and 1.3~mrad convergence semiangle, with a step size of 10~\AA{} and a scan size of 200x50 probe positions. 

To obtain an independent measurement of the sample strain, we also acquired an electron energy loss spectrum (EELS) dataset from the same region of the sample. Analysis of the EELS data showed the average thickness to be approximately one inelastic mean free path, corresponding to an estimated thickness of 110~nm. Chemical analysis showed the Si region to be pure Si, and the SiGe alloy region to have an average composition of 18\% Ge. From this chemical analysis we can derive the expected strain in the SiGe layers.

First, we use Vegard's law, which posits that the strain depends linearly on the composition $x_{\rm{Si}}$ \cite{vegard1921konstitution}.  The Si$_{0.82}$Ge$_{0.18}$ layers have a larger lattice constant, and thus will expand relative to the Si layers in the $x$ direction. Because the multilayers are epitaxial, the Si$_{0.82}$Ge$_{0.18}$ layers are compressed in the multilayer interfacial plane in two directions, which will lead to an additional expansion given by the Poisson's ratio multiplied by two. The overall strain profile can therefore be estimated as
\begin{equation}
    \epsilon_{\rm{xx}} 
    = 
    \left(
        \frac{a_{\rm{Ge}}}{a_{\rm{Si}}}-1
    \right)
    (1 - x_{\rm{Si}})
    (1 + 2 \nu),
\end{equation}
which is plotted in Fig.~\ref{Fig:StrainMap_exp}i, using literature values for the cubic lattice constants of Si and Ge of $a_{\rm{Si}} = 5.54$ and $a_{\rm{Ge}} = 5.66$ \AA{}, respectively \cite{johnson1954some}, and for the Poisson's ratio $\nu$ of Si and Ge of approximately 0.275 in the (001) direction \cite{wortman1965young}.

\section*{Data and Code Availability}

 Codes related to FCU-Net model, data preprocessing and augmentation can be found in \crystalFD{} repository and are available as open source package. Distributed Hyperparameter tuning pipeline using rayTune can be found \href{https://github.com/AI-ML-4DSTEM/4D-OPTIMIZE/tree/nersc_ray}{here}.
 
 Disk detection using AI/ML (FCU-Net) is implemented as a new functionality in \pyFDSTEM{} 0.12.x. The simulated and experimental strain measurements performed in this paper and the required 4D-STEM dataset are available as tutorial notebooks and can be accessed \href{https://github.com/py4dstem/py4DSTEM_tutorials/tree/main/notebooks/strain_aiml}{here}.
 
 Dynamical diffraction library generation tool and the simulated training dataset are available upon reasonable request.

\section*{Acknowledgements}

This work was primarily funded by the US Department of Energy in the program ``4D Camera Distillery: From Massive Electron Microscopy Scattering Data to Useful Information with AI/ML.''  MKYC and CO each acknowledges support of a US Department of Energy Early Career Research Award. JC acknowledges support from the Presidential Early Career Award for Scientists and Engineers (PECASE) through the U.S. Department of Energy. BHS and \pyFDSTEM{} development are supported by the Toyota Research Institute. SEZ was supported by the National Science Foundation under STROBE Grant no. DMR 1548924. Work at the Molecular Foundry was supported by the Office of Science, Office of Basic Energy Sciences, of the US Department of Energy under Contract No. DE-AC02-05CH11231. Use of the Center for Nanoscale Materials, an Office of Science user facility, was supported by the U.S. Department of Energy, Office of Science, Office of Basic Energy Sciences, under Contract No. DE-AC02-06CH11357. This research used resources of the National Energy Research Scientific Computing Center, a DOE Office of Science User Facility supported by the Office of Science of the U.S. Department of Energy under Contract No. DE-AC02-05CH11231. We acknowledge S. Kim, and J. Carlstroem for assistance with sample preparation.  We also acknowledge donation of GPU resources by NVIDIA.

\section*{References}
\bibliography{refs}
\bibliographystyle{apsrev4-1}

\end{document}